\documentclass[twocolumn,prl,showpacs,groupedaddress]{revtex4}
\usepackage{graphicx}

\begin{document}

\title{Loschmidt echo and Berry phase around degeneracies in nonlinear systems}
\author{X. X. Yi, X. L. Huang, and W. Wang}
\affiliation{School of Physics and Optoelectronic Technology,
              Dalian University of Technology, Dalian 116024, China}

\date{\today}

\begin{abstract}
We study the Loschmidt echo and Berry phase in a nonlinear system
transported around a double  and triple degeneracy. The nonlinearity
of the system makes the Berry phase different from that in linear
systems. We propose a witness of nonlinearity  for the nonlinear
system and show its dependence on the parameters of the system,
taking the standard Landau-Zener model as an example. We calculate
the Loschmidt echo(LE) or quantum fidelity in the quantum dynamics
under perturbations around the degeneracy, and establish a
connection between the LE and the witness of nonlinearity. These
phases and Loschmidt echo can be observed with current experimental
technology in a nonlinear resonator.
\end{abstract}

 \maketitle

Since Berry's introduction\cite{berry84} of the adiabatic geometric
phase, a large number of article have appeared on the theoretical
foundations, physical applications, and experimental manifestations
of geometric phases \cite{shapere89,bohm03}. Although there are by
now hundreds papers on geometric phases, there have no studies on
this subject in nonlinear systems  until seminal works\cite{liu03,
wu05}, due to the lack of orthogonality of its Hamiltonian
eigenstates and the linear superposition principle.

Singularities in mathematics  are often associated with specific
effects in physics. For instance, topological singularities such as
diabolic points are associated with a specific phase behavior of the
wave function\cite{berry85,sachdev99}. Exceptional points as the
other type of topological singularity are associated with level
repulsions and can affect Rabi oscillations \cite{dietz07}.  For a
linear system, it has been shown that the eigenfunction of a real
Hamiltonian can acquire a $\pi$ geometric phase(a sign change to the
wave function) when it is transformed around a certain type of
degeneracy. This sign change was found in the late 19th century,
however its significance to physics was not realized until
Longuet-Higgins and coworkers shown its existence in molecular
physics\cite{longuet58}. The latter insight leads to the notion of
the molecular Aharonov-Bohm effect\cite{mead79} that has attracted a
lot of attention both experimentally and theoretically in recent
years\cite{bush98, kendrick97, sjoqvist02}. For  nonlinear systems,
we may have additional eigenvalues and eigenvectors more than its
dimension of Hilbert space. The degeneracy in this system is also
different from that in linear systems, for example, the
eigenfunctions at the degeneracy point may be not orthogonal to each
other in general. This makes the Berry phase around a degeneracy
 in nonlinear systems different from that in linear systems.

In this Letter, we shall study the Berry phase in the vicinity of
double degeneracies in nonlinear systems. A general expression for
the Berry phase will be given and discussed. The results show that
the Berry phase around the degeneracy significantly depends on the
nonlinearity characterized by an overlap of the degenerate
eigenvectors at the degeneracy point.  This overlap will be defined
as a witness of nonlinearity, which is found to be related to the
Loschmidt echo in the quantum dynamics around the degeneracy. As an
example, we calculate the Berry phase and the witness of
nonlinearity around the double degeneracy in the standard
Landau-Zener model.  An extension from double to triple degeneracy
is also presented and discussed.

We start by recalling the calculation of the Berry phase in a
general system (linear or nonlinear) with Hamiltonian
$H(X)$\cite{wu05}, where $X=(X_1,X_2,...,X_m)$ is a vector parameter
that the system depends on. For a quantum system starting at an
eigenstate $|\Psi_n(X)\rangle$ defined by
$H(X)|\Psi_n(X)\rangle=E_n(X)|\Psi_n(X)\rangle$, its state at time
$t$ may be written as
$|\Psi(t)\rangle=e^{-i\beta(t)}|\Psi_n(X)\rangle$ when $X$ changes
adiabatically. The Berry phase in this case is
\begin{equation}
\gamma_n=i\oint_c \frac{\langle \Psi_n(X)|\partial /\partial
X|\Psi_n(X)\rangle}{\langle \Psi_n(X)|\Psi_n(X)\rangle}dX.
\label{bdef}
\end{equation}
Eq.(\ref{bdef}) is valid for both linear and nonlinear quantum
systems as long as the system is initially in an eigenstate of the
system Hamiltonian. Eq.(\ref{bdef}) simplifies after applying the
standard normalization $\langle \Psi_n(X)|\Psi_n(X)\rangle=1.$
Consider a double degeneracy\cite{note1} $E_n=E_{n+1}$, we denote
$X_d$ the degenerate point in the parameter space, i.e.,
$E_n(X_d)=E_{n+1}(X_d)=E_d.$ The eigenvectors at this point
$|\Psi_n^d\rangle=|\Psi_n(X_d)\rangle$ and
$|\Psi_{n+1}^d\rangle=|\Psi_{n+1}(X_d)\rangle$ are chosen to be
normalized ($\langle \Psi_n^d|\Psi_n^d\rangle=1=\langle
\Psi_{n+1}^d|\Psi_{n+1}^d\rangle$). The orthogonality condition
$\langle\Psi_n^d|\Psi_{n+1}^d\rangle =0$ holds for linear systems,
but it  is not satisfied for nonlinear systems in general. As we
shall show, this  makes the Berry phase around the double degeneracy
different for  linear  and nonlinear systems.

Consider a small circular loop around the double degeneracy point
$X_d$, $C=\{ X(t)=X_d+\delta X(t)\}$ with $\delta X(0)=\delta X(T).$
Here $t \in [ 0,T]$ represents the time and $T$ stands for the
duration of the cyclic evolution. Suppose that the perturbation
theory holds around the double degeneracy point $X_d,$ the
eigenvectors and eigenvalues at point $X(t)$ take the asymptotic
form, $
|\Psi_n(X(t))\rangle=\cos\frac{\theta}{2}|\Psi_n^d\rangle+\sin\frac{\theta}{2}
e^{-i\phi}|\Psi_{n+1}^d\rangle,$ $ |\Psi_{n+1}(X(t))\rangle=\sin
\frac{\theta}{2}|\Psi_n^d\rangle-\cos\frac{\theta}{2}e^{-i\phi}
|\Psi_{n+1}^d\rangle,$ $ E_j(X(t)) = E_d+\delta E_j, j=n, n+1, $
where $\delta E_j$ are eigenvalues of the following matrix,
\begin{equation}
\delta H=\left( \matrix{ \delta H_{n,n} & \delta H_{n,n+1} \cr
 \delta H_{n+1,n} & \delta H_{n+1,n+1} \cr } \right).\label{ha2}
\end{equation}
Here $\delta H_{\alpha, \beta}=\langle \Psi_{\alpha}^d|\sum_{k=1}^m
\frac{\partial H}{\partial X_k}|_{X=X_d}\delta
X_k(t)|\Psi_{\beta}^d\rangle, \ \ \ \alpha, \beta =n, n+1.$ After
some simple manipulations, we find that $ \delta
E_{n,n+1}=\frac{\delta H_{n,n}+\delta H_{n+1,n+1}}{2} \pm
\sqrt{\frac 1 4 (\delta H_{n,n}-\delta H_{n+1,n+1})^2+|\delta
H_{n,n+1}|^2},$ and $ \cos \theta =\frac{\delta H_{n,n}-\delta
H_{n+1,n+1}}{\sqrt{(\delta H_{n,n}-\delta H_{n+1,n+1})^2+4|\delta
H_{n,n+1}|^2}},$ $ \phi=\mbox{Arg}(\delta H_{n,n+1}).$ By using Eq.
(\ref{bdef}), we arrive at an asymptotic expression for the Berry
phase,
\begin{equation}
\gamma_{n,n+1}=\frac 1 2 \oint_c \frac{(1\mp
\cos\theta)}{\Delta_{\pm}} d\phi \pm \frac 1 2 \oint_c
\frac{\sin\theta |\langle \Psi_n^d|\Psi_{n+1}^d\rangle|}{\Delta_{\pm}}d\phi,\nonumber\\
\label{bp1}
\end{equation}
Here $\Delta_\pm=1\pm
\sin\theta|\langle\Psi_n^d|\Psi_{n+1}^d\rangle|,$ and $\theta$ was
assumed to be $\phi$-independent.  These phases simplify, $
\gamma_n=\frac \pi \Delta_+(1-\cos\theta)+\frac \pi \Delta_+
\sin\theta |\langle \Psi_n^d|\Psi_{n+1}^d\rangle |,$ $
\gamma_{n+1}=\frac \pi \Delta_- (1+\cos\theta)-\frac \pi \Delta_-
\sin\theta |\langle \Psi_n^d|\Psi_{n+1}^d\rangle |,$ when $|\langle
\Psi_n^d|\Psi_{n+1}^d\rangle|$ is independent of $\phi.$ For linear
systems, the eigenvectors $|\Psi_n^d\rangle$ and
$|\Psi_{n+1}^d\rangle$ are orthogonal, leading the Berry phases to
$\gamma_n=\frac{\Omega_c}{2}$ and $\gamma_{n+1}=-\frac{\Omega_c}{2}$
where $\Omega_c$ is the solid angle subtended by $C$ on the unit
sphere in the space. For a real Hamiltonian that describes
reversible systems, $|\Psi_n^d\rangle$ and $|\Psi_{n+1}^d\rangle$
can be chosen to be real such that the cycle  $C$ around the
degeneracy point $X_d$ lies in a plane\cite{mailybaev05}. As a
consequence we have $\gamma_n=\gamma_{n+1}=\pi$ if $C$ makes a
single turn around the degeneracy point, while the Berry phase would
be zero if the degeneracy point lies outside the
cycle\cite{berry84}.

For nonlinear systems, however, the eigenvectors $|\Psi_n^d\rangle$
and $|\Psi_{n+1}^d\rangle$ are not orthogonal in general. Assuming
$|\langle \Psi_n^d|\Psi_{n+1}^d\rangle |$ to be independent of
$\phi$ and very small ($|\langle\Psi_n^d|\Psi_{n+1}^d\rangle|\ll
1$), the Berry phases in this case reduce to, $ \gamma_{n,n+1}=\pm
\frac{\Omega_c}{2}+(\frac{\Omega_{c^{\prime}}}{4} +\frac \pi
2)|\langle\Psi_n^d|\Psi_{n+1}^d\rangle|, $ where
$\Omega_{c^{\prime}}$ was defined as $
\Omega_{c^{\prime}}=-\oint_c[1+\cos(\frac{\pi}{2}-2\theta)]d\phi$
that stands for the solid angle with
$\theta^{\prime}=(\frac{\pi}{2}-2\theta)$ instead of $\theta$ in
$\Omega_c.$ For $|\langle\Psi_n^d|\Psi_{n+1}^d\rangle| \rightarrow
1$, the Berry phases reduce to $ \gamma_{n+1,n}=\frac 1 2 \oint(1\pm
\frac{\cos\theta}{1\mp\sin\theta})d\phi. $ These analysis show that
the Berry phase around the degeneracy point significantly depends on
the overlap $|\langle\Psi_n^d|\Psi_{n+1}^d\rangle|$ and this overlap
might be related to the nonlinearity of the system. Indeed, as we
shall demonstrate in the next paragraph, the overlap may be chosen
as a witness of nonlinearity. This witness of nonlinearity can be
observed through the Loschmidt echo as follows. The LE was defined
as the overlap between two states that evolve from the same initial
wave function $|\Psi_0\rangle$ under two slightly different
Hamiltonians $H(X_d)$ and $H(X_d)+\delta H,$ respectively,
$L(t)=|\langle\Psi_0|U^{\dagger}(H+\delta H)U(H)|\Psi_0\rangle|^2,$
where $U(h)$ $(h=H(X_d), H(X_d)+\delta H)$ stands for the time
evolution operator corresponding to Hamiltonian $h.$ Under the
adiabatic evolution, the LE reads,
\begin{equation}
L(t)=\frac{1}{\Delta_+}|\cos
\frac{\theta}{2}+\sin\frac{\theta}{2}|\langle
\Psi_n^d|\Psi_{n+1}^d\rangle ||^2,\label{le}
\end{equation} where
$|\Psi_0\rangle=|\Psi_n^d\rangle$ was assumed. This result indicates
that the LE, which quantifies the stability of the quantum dynamics
of a system against perturbations is closely connected to  $\langle
\Psi_n^d|\Psi_{n+1}^d\rangle$, i.e., the overlap can be a good
witness of nonlinearity.  For linear systems, $\langle
\Psi_n^d|\Psi_{n+1}^d\rangle =0$ leading to
$L(t)=\cos^2\frac{\theta}{2}.$ When the Hamiltonian $H(X_d)+\delta
H$ drives the system to follow the circular loop that lies in the
equator of the sphere $(r,\theta,\phi)$, where $r=\frac 1 2 (\delta
E_{n+1}-\delta E_n)$, we have $\cos\theta =0$, then the LE becomes
$L(t)=\frac 1 2.$ In this case,  the Berry phases reduce to
$\gamma_{n+1}=\gamma_n=\pi.$
 $\gamma_{n,n+1}$ in Eq.(\ref{bp1}) can
reproduce the Berry phase in linear systems even with nonzero
overlap. This happens when $|\delta H_{n,n+1}| \ll |\delta
H_{n,n}-\delta H_{n+1,n+1}|$ reminiscent of the adiabatic condition
in linear systems. The Loschmidt echo in this situation takes
$L(t)=|\langle \Psi_n^d|\Psi_{n+1}^d\rangle|^2$ with $\delta
H_{n,n}<\delta H_{n+1,n+1}$, and $L(t)=1$ with $\delta
H_{n,n}>\delta H_{n+1,n+1}.$ This feature clearly bridges the
Loschmidt echo and the witness of nonlinearity and makes the witness
of nonlinearity experimentally observable. The Loschmidt echo may
decay in quantum systems whose classical counterparts have strong
chaos with exponential instability, this decay  has been weel
studied in\cite{jalabert01, jacquod02}. Eq.(\ref{le}) sheds light on
this instability form a new point of view, i.e., the overlap
$|\langle\Psi_n^d|\Psi_{n+1}^d\rangle|.$ We see that $L(t)$ is
independent of perturbations $\delta E_{n,n+1}$, therefore the
Loschmidt decay (if any) falls into the Lyapunov regime, providing
us an other way to study quantum chaos.

\begin{figure}
\includegraphics*[width=0.9\columnwidth,
height=0.7\columnwidth]{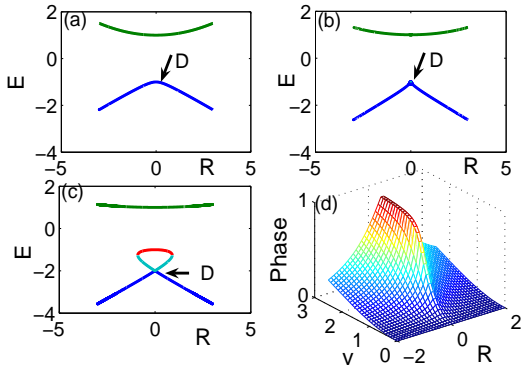} \caption{(color online) Energy
levels and the Berry phase as a function of $R$ and $v$.
Fig.\ref{fig1}-(a),(b) and (c) are plotted for the energy level with
(a) $v=2$, (b) $v=1$, and (c) $v=0.5$. The phase in (d) was computed
in units of $\pi$. The other parameter chosen is $c=1.$ }
\label{fig1}
\end{figure}

\begin{figure}
\includegraphics*[width=0.8\columnwidth,
height=0.6\columnwidth]{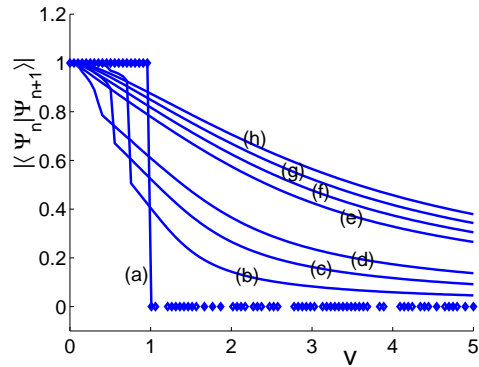} \caption{(color online)
Overlap of the two degenerate wavefunction. $c=1$ was set for this
plot. (a)-(h) correspond to $R=0, 0.2,0.4,0.8,1.4,1.6,1.8,$ and
$2.0$, respectively. } \label{fig2}
\end{figure}

As an example, we demonstrate the Berry phase $\gamma_n$ with a
nonlinear two-level model as,
$
i\frac{\partial}{\partial t}\left(\matrix{\psi_1\cr
\psi_2}\right)=\left( \matrix{ \frac R 2+\frac c 2 m & \frac v 2
e^{i\phi}\cr
  \frac v 2 e^{-i\phi} & -\frac R 2-\frac c 2 m \cr } \right)\left(\matrix{\psi_1\cr
\psi_2}\right),\label{exa} $ where $m=|\psi_2|^2-|\psi_1|^2$,
$\psi_1$ and $\psi_2$ are the probability amplitudes. $v e^{i\phi}$
is the complex coupling between the two levels, $c$ stands for the
nonlinear parameter that characterizes the dependence of the level
energy on the populations. $R$ is the level bias. This model can be
used to describe the Josephson effect of Bose-Einstein condensates
in a double-well potential\cite{leggett01,milburn97}. The complex
coupling can be realized in experiment with current
technology\cite{denschlag00}. To study the Berry phase, we need to
find all the eigenstates $|\Psi_n\rangle$ of the nonlinear
Hamiltonian $H=\left( \matrix{ \frac R 2+\frac c 2 m & \frac v 2
e^{i\phi}\cr
  \frac v 2 e^{-i\phi} & -\frac R 2-\frac c 2 m \cr } \right).$ By
Eq.(\ref{bdef}), we obtain the Berry phase
\begin{equation}
\gamma=\pi(1-\sqrt{1-\frac{v^2}{4E^2}}),\label{bp}
\end{equation}
where $E$ is one of the real roots of equation, $E^4+cE^3+\frac 1
4(c^2-v^2-R^2)E^2-\frac{v^2c}{4}E-\frac{v^2c^2}{16}=0.$ To derive
Eq.(\ref{bp}), we restricted the system to follow a path with fixed
$R, c$ and $v$, i.e., only $\phi$ is allowed to change. It may have
at most 4 real roots, indicating more than two eigenstates can exist
in that system(see Fig. \ref{fig1}-(a), (b) and (c)).
In order to involve  the Berry phase around the degeneracy point $D$
$(R=0)$, where the first and second eigenvalues touch, we choose the
lowest eigenvalue as the $E$ in Eq.(\ref{bp}), we plot the Berry
phase as a function of $R$ and $v$ in figure \ref{fig1}-(d). The
nonlinearity parameter $c=1$ was set for this plot. For weak
nonlinearity ($v\gg c$), the Berry phase is $\pi$ at $R=0,$ while it
is zero for strong nonlinearity($v\ll c$). As shown in \cite{wu00}
$c=v$ (Fig.\ref{fig1}-(b)) is a critical point for the system to
have more eigenvalues. For $c<v$ (Fig.\ref{fig1}-(a)) there are two
eigenvalues while there can be four eigenvalues when $c>v$
(Fig.\ref{fig1}-(c)). We can find this critical point in
Fig.\ref{fig2}, where a step change among the line $R=0$ occurs in
the overlap $|\langle\Psi_n|\Psi_{n+1}\rangle|$. Two observations
can be made from Fig. \ref{fig2}. (1) As $v/c$ increases, the
overlap tends to zero, indicating the system changes from nonlinear
to linear one. (2) A step change in the overlap occurs at the
degenerate point $R=0$, it becomes unclear with $R$ far from the
degeneracy point. These features makes the overlap a good witness
for the nonlinearity and critical point.

We now extend the presentation to the case of triple degeneracy.
Consider a Hamiltonian that removes the threefold degeneracy
$E_{n-1}=E_n=E_{n+1},$
\begin{equation}
\delta H=\left( \matrix{ \delta H_{n-1,n-1} & \delta H_{n-1,n} &
\delta H_{n-1,n+1} \cr
 \delta H_{n,n-1} & 0 & 0 \cr
\delta H_{n+1,n-1} & 0 & 0\cr
  } \right),\label{ha3}
\end{equation}
where $H_{\alpha,\beta}, \alpha, \beta= n-1, n, n+1$ is the same as
in Eq.(\ref{ha2}) but here the degeneracy are threefold, and the
degenerate wavefunction $|\Psi_{n-1}\rangle$, $|\Psi_{n}\rangle$ and
$|\Psi_{n+1}\rangle$ are chosen to be real.
 This approximation consists of neglecting transition
amplitudes between the three degenerate states and other states as
well as the transitions between  $n$th and $(n+1)$th degenerate
states. $\delta H_{n+1,n+1}$ and $\delta H_{n,n}$ are also assumed
to be equal. This kind of Hamiltonian can be realized by treating
the threefold degenerate states as a three-level $\Lambda$-system,
labeling one of degenerate states as $|e\rangle$ and the other two
as $|a\rangle$ and $|g\rangle$. The state $|e\rangle$ is coupled to
state $|a\rangle$ and $|g\rangle$ with coupling constant $\delta
H_{n-1,n}$ and $\delta H_{n-1,n+1}$, respectively. The
$\Lambda$-type system has been intensively studied in quantum optics
and recently it was used to discuss the adiabatic condition for
nonlinear systems \cite{pu07}. Defining $\Omega^2=(\delta
H_{n-1,n-1})^2+4(\delta H_{n-1,n})^2+4(\delta H_{n-1,n+1})^2,$
$\sin\theta \cos\phi=\frac 2 \Omega \delta H_{n-1,n},$ and
$\sin\theta \sin\phi=\frac 2 \Omega \delta H_{n-1,n+1},$ we can
write the eigenstates and eigenvalues for $\delta H$ as,
\begin{eqnarray}
|\Psi_{n-1}\rangle &=& -\cos\frac \theta
2(\sin\phi|\Psi_{n+1}^d\rangle +
\cos\phi|\Psi_{n}^d\rangle)+\sin\frac \theta
2|\Psi_{n-1}^d\rangle,\nonumber\\
|\Psi_{n}\rangle &=& \cos\phi|\Psi_{n+1}^d\rangle -
\sin\phi|\Psi_{n}^d\rangle,\nonumber\\
|\Psi_{n+1}\rangle &=& \sin\frac \theta
2(\sin\phi|\Psi_{n+1}^d\rangle +
\cos\phi|\Psi_{n}^d\rangle)+\cos\frac \theta
2|\Psi_{n-1}^d\rangle.\nonumber\\ \label{statetri}
\end{eqnarray}
Here $\theta$ can varies from $0$ to $2\pi$, the triple degeneracy
happens at $\Omega=0$( the origin of the sphere
($r=\Omega,\theta,\phi$)). Eq.(\ref{statetri}) shows that
$|\Psi_n(\theta+\pi)\rangle=|\Psi_n(\theta)\rangle,$
$|\Psi_{n-1}(\theta+\pi)\rangle=|\Psi_{n+1}(\theta)\rangle,$ and
$|\Psi_{n+1}(\theta+\pi)\rangle=-|\Psi_{n-1}(\theta)\rangle,$
implying $|\Psi_n(\theta+2\pi)\rangle=|\Psi_n(\theta)\rangle,$
$|\Psi_{n-1}(\theta+2\pi)\rangle=-|\Psi_{n-1}(\theta)\rangle,$ and
$|\Psi_{n+1}(\theta+2\pi)\rangle=-|\Psi_{n+1}(\theta)\rangle.$ In
fact, the latter relation can be found straightforwardly from
Eq.(\ref{statetri}). This completes the picture when the circular
path lies in the great circle of the sphere. In the case where the
path lies in the equator of the sphere, namely $\theta=\frac \pi 2$,
Eq.(\ref{statetri}) follows that
$|\Psi_n(\phi+2\pi)\rangle=|\Psi_n(\phi)\rangle,$
$|\Psi_{n-1}(\phi+2\pi)\rangle=|\Psi_{n-1}(\phi)\rangle,$ and
$|\Psi_{n+1}(\phi+2\pi)\rangle=|\Psi_{n+1}(\phi)\rangle.$ So for the
triple degeneracy, we have two allowed adiabatic sign changes around
the degeneracy, which were listed below in table \ref{tab1}. It is
worth pointing our that the Berry phases presented in this case are
all for loops which {\it enclose} the point of degeneracy. By {\it
enclose} we mean the loop can not be smoothly deformed to avoid
surrounding the degeneracy point in the parameter space. This is
different from that for the double degeneracy, where the Berry phase
is calculated for a general circular loop around the degenerate
point.

\begin{table*}
\renewcommand{\arraystretch}{1.3}
\caption{Allowed adiabatic sign changes around the triple degeneracy
in the non-linear system.}
\begin{tabular}{|p{2cm}|p{ 3.8cm}|p{3.8cm}|p{3.8cm}|}
\hline &$\langle \Psi_{n-1}(\alpha)|\Psi_{n-1}(\alpha+2\pi)\rangle$
&$\langle \Psi_{n}(\alpha)|\Psi_{n}(\alpha+2\pi)\rangle$
&$\langle \Psi_{n+1}(\alpha)|\Psi_{n+1}(\alpha+2\pi)\rangle$\\
\hline
$\alpha=\phi$ & +1 & +1 & +1   \\
\hline
$\alpha=\theta$ & -1 & +1 & -1 \\
\hline
\end{tabular}
\label{tab1}
\end{table*}
\begin{figure}
\includegraphics*[width=0.7\columnwidth,
height=0.5\columnwidth]{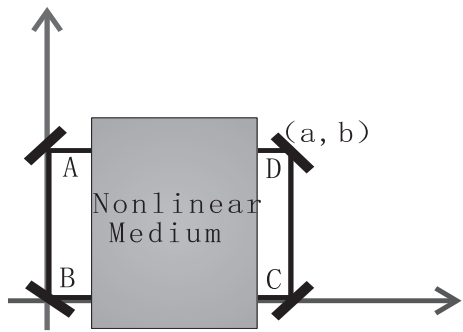} \caption{ Sketched setup of
the rectangular resonator. $(a,b)$ is the position of mirror D where
three levels are degenerate.} \label{fig3}
\end{figure}
The presented prediction is within touch of recent experimental
technology. The observation of the Berry phase around a triple
degeneracy is available by the  setup in \cite{lauber94} but with
Kerr nonlinearity. Due to the nonlinear interaction, the triple
degeneracy point may change, however it can be  relocated by
measuring the spectra of the rectangular resonator that is a
function of the shape of the resonator (representing by $(a,b)$ as
shown in Fig.\ref{fig3}).  The geometric phase effect then can be
measured  as Lauber {\it et al.} did in Ref. \cite{lauber94} as
follows. (1) After finding the degeneracy point, shift the position
of the mirror at one of its corners (say D) around $(a,b)$. (2)
Measure standing wave patterns via the reflected microwave intensity
by the technology in \cite{stein92}. The Berry phase effect can be
found by comparing those patterns at the start and end point. The
Berry phase around the double degeneracy can observed in the same
manner, by finding a double degeneracy instead of the triple
degeneracy in the spectra of triangular resonators. Alternatively,
Bose-Einstein condensates in a double-well potential may meet the
requirement to observe the Berry phase around a double degeneracy,
serving as the nonlinear system.

In conclusion, we have studied the Loschmidt echo and Berry phase
around a degeneracy in nonlinear systems. Two degeneracies, a double
degeneracy and a triple degeneracy, are considered. We have found
that the Berry phases are different from that in linear systems due
to nonorthogonality of the degenerate wave functions at the
degenerate point. We have also found that the overlap of the
degenerate wavefunctions may serve as a witness of nonlinearity of
nonlinear systems. A connection between this witness and the
Loschmidt echo has been established. Allowed adiabatic sign changes
around the degenerate point are presented and discussed. The above
analysis is based on the first order perturbation theory, which
fails in the presence of additional satellite degeneracy near the
main degeneracy\cite{child99}, this problem can be solved in linear
systems  by taking the second order perturbation into
account\cite{pistolesi00}.

This work was supported by   NSF of China under Grant  No. 60578014
and 10775023, and
National Basic Research Program of China.\\


\begin{references}
\bibitem{note1} The present analysis is for level-crossing, it is
easy to be generalized for avoided level-crossing.
\end{references}
\end{document}